\documentclass[
    ,final            % use final for the camera ready runs
    ,nomathfonts
  ]
  {aipproc}

\layoutstyle{8x11single}

\begin{document}

\title{Probing the gateway to superheavy nuclei in cranked
relativistic Hartree-Bogoliubov theory}

\author{A.\ V.\ Afanasjev\footnote{on leave of absence
from the Laboratory of Radiation Physics, Institute of Solid State
Physics, University of Latvia, LV 2169 Salaspils, Miera str. 31, Latvia}\,\,}{
address={Physics Division, Argonne National Laboratory,
Argonne, IL 60439, USA},
altaddress={Department of Physics, University of Notre Dame,
Notre Dame, Indiana 46556, USA}
}

\author{T.\ L.\ Khoo}{
address={Physics Division, Argonne National Laboratory,
Argonne, IL 60439, USA}
}

\author{S.\ Frauendorf}{
address={Department of Physics, University of Notre Dame,
Notre Dame, Indiana 46556, USA}
}

\author{G.\ A.\ Lalazissis}{
address={Department of Theoretical Physics, Aristotle University of
Thessaloniki, GR54124, Thessaloniki, Greece}
}

\author{I.\ Ahmad}{
address={Physics Division, Argonne National Laboratory,
Argonne, IL 60439, USA}
}

\begin{abstract}
The cranked relativistic Hartree+Bogoliubov theory has been applied for
a systematic study of the nuclei around $^{254}$No, the heaviest nuclei
for which detailed spectroscopic data are available. The deformation,
rotational response, pairing correlations, quasi-particle and other
properties of these nuclei have been studied with different relativistic
mean field  (RMF) parametrizations. For the first time, the quasi-particle 
spectra of odd deformed nuclei have been calculated in a fully self-consistent
way within the framework of the RMF theory. The energies of the 
spherical subshells, from which active deformed states of these nuclei
emerge, are described with an accuracy better than 0.5 MeV for most of
the subshells with the NL1 and NL3 parametrizations. However, for a few
subshells the discrepancy reach 0.7-1.0 MeV. The implications of these
results for the study of superheavy nuclei are discussed.
\end{abstract}

\maketitle

%%%%%%%%%%%%%%%%%%%%%%%%%%%%%%%%%%%%%%%%%%%%
%% MAINMATTER
%%%%%%%%%%%%%%%%%%%%%%%%%%%%%%%%%%%%%%%%%%%%

%################################
\section{Introduction}
%################################

 The possible existence of shell-stabilized superheavy nuclei, predicted 
with realistic nuclear potentials  \cite{SGK.66,C.67,M.67} and the 
macroscopic-microscopic (MM) method \cite{Nilsson68,Nilsson69,MG.69}, 
has been a driving force behind experimental and theoretical efforts 
to investigate the superheavy nuclei. These investigations pose a number 
of experimental and theoretical challenges. On the theoretical side,
no consensus has been achieved on the question of what are the 
magic shell gaps in superheavy nuclei. The situation is illustrated in 
Table \ref{predictions}, where the predictions of different models are 
summarized.

  The accuracy of predictions of spherical shell closures depends sensitively 
on the accuracy of describing the single-particle energies, which becomes 
especially important for superheavy nuclei, where the level density is very high. 
Variations in single-particle energy of $1-1.5$ MeV yield spherical shell gaps 
at different particle numbers, which restricts the reliability in extrapolating 
to an unknown region. 

  The MM method describes the single-particle energies rather well in known regions. 
This is due to the fact that the experimental data on single-particle states are 
used directly in the parametrization of  the single-particle potential. However, 
the extrapolation of the single-particle potential may be much less reliable 
since it is not determined self-consistently. For example, microscopic models 
predict that the appearance of shell closures in superheavy nuclei is 
influenced by a central depression of the nuclear density distribution 
\cite{BRRMG.99,Detal99}. This effect is not treated in a self-consistent way 
in current MM models.

 Although the nucleonic potential is defined in self-consistent approaches,
such as Skyrme Hartree-Fock (SHF) and relativistic mean field (RMF) theory, 
in a fully self-consistent way, this does not guarantee that single-particle 
degrees of freedom are accurately described. This is especially true because
the parameters of the Skyrme forces and RMF Lagrangians were fitted
mostly to bulk properties, and the accuracy of the description of the 
single-particle energies is poorly known. Compared with the MM method, 
self-consistent calculations have been confronted with experiment to a lesser 
degree and for a smaller number of physical observables (mainly binding 
energies and quantities related to their derivatives). For many parametrizations, 
even the reliability of describing conventional nuclei is poorly known.

%%%%%%%%%%%%%%%%%%%%%%%%%%%%%%%%%%%%%%%%%%%%
\begin{table}
\begin{tabular}{lrrrr}
\hline
    \tablehead{1}{r}{b}{Method}
  & \tablehead{1}{r}{b}{Proton\\shell gap}
  & \tablehead{1}{r}{b}{Neutron\\shell gap}
  & \tablehead{1}{r}{b}{Potential\\ in the MM method}
  & \tablehead{1}{r}{b}{References}   \\
\hline
MM     & 114     & 184     & Nilsson     & \cite{Nilsson69}   \\
       &         &         & Woods-Saxon & \cite{Nilsson68,PS.91,CDHMN.96} \\
       &         &         & folded Yukawa & \cite{MN.94} \\      
Skyrme & 126\tablenote{Only the values appearing in most
of the parametrizations are quoted. Some Skyrme forces 
indicate $Z=114$ (SkI4) and $Z=120$ (SkI3) as proton shell 
closures, while some (for example, SkP) predict no doubly 
magic superheavy nuclei at all.}     
       & 184     &       &  \cite{CDHMN.96,RBBSRMG.97,BRRMG.99}  \\
Gogny  & 120/126 & 172/184 &       &  \cite{BBDGD.01}  \\
RMF    & 120\tablenote{The NLSH (NLRA1) parametrizations of 
the RMF Lagrangian give (also) $Z=114$ and $N=184$ as shell closures \cite{LSRG.96,NL-RA1}, 
but since they give a poor description of quasiparticle 
spectra in the deformed $A\sim 250$ mass region, we consider these 
predictions less reliable than those obtained with other RMF sets.}
       &  172    &       &  \cite{RBBSRMG.97,BRRMG.99}  \\
\hline
\end{tabular}
\caption{Predicted magic spherical shell gaps for superheavy nuclei.}
\label{predictions}
\end{table}
%%%%%%%%%%%%%%%%%%%%%%%%%%%%%%%%%%%%%%%%%%%%%%%%%%%%%%%%%%%%%%%%%%%%%%%%%%%%%%%%%%

 In order to fill this gap in our knowledge, the cranked relativistic 
Hartree+Bogoliubov (CRHB) theory  \cite{A190,CRHB} has been applied for 
a systematic study of the nuclei around $^{254}$No, the heaviest elements 
for which detailed spectroscopic data are available. The 
deformations, rotational response, pair correlations, quasiparticle spectra, 
shell structure and the two-nucleon separation energies have been studied. The 
goal was to see how well the theory describes the experimental data and how
this description depends on the RMF parametrization. The details of this
study will be reported in a forthcoming manuscript \cite{AKFAL.02}. 

In the present contribution, we mainly concentrate on the results having 
implications for the study of superheavy nuclei. Particular attention is 
paid to the comparison of experimental and calculated quasiparticle spectra 
in deformed nuclei and, based on that, how one can estimate the accuracy 
of the calculated energies of spherical subshells. One can then (i) judge  
which RMF parametrizations provide best description of single-particle 
energies and thus are best suited for the study of superheavy nuclei and 
(ii) assess how the RMF predictions for superheavy nuclei are modified if 
empirical shifts for the energies of spherical subshells, deduced from the 
study of deformed nuclei, are taken into account.

%%%%%%%%%%%%%%%%%%%%%%%%%%%%%%%%%%%%%%%%%%%%%%%%%%%%%%%%%%%%%%%%%%%%
\section{Shell structure in the deformed $A\sim 250$ mass region.}
%%%%%%%%%%%%%%%%%%%%%%%%%%%%%%%%%%%%%%%%%%%%%%%%%%%%%%%%%%%%%%%%%%%%

 The stability of the superheavy elements is due to a 'shell gap', 
i.\ e.\ a region of low level density in the single-particle spectrum. 
The quantity $\delta_{2n}(Z,N)$ related to the derivative of the 
separation energy is a sensitive indicator of the localization
of the shell gaps. For the neutrons (and similarly for the protons) 
it is defined as
\begin{eqnarray}
\delta_{2n}(Z,N)=S_{2n}(Z,N)-S_{2n}(Z,N+2)= \nonumber \\
=-B(Z,N-2)+2B(Z,N)-B(Z,N+2)
\label{2n-shell-gap}
\end{eqnarray}
where $B(N,Z)$ is the binding energy. 

%%%%%%%%%%%%%%%%%%%%%%%%%%%%%%%%%%%%%%%%%%%%%%%%%%%%%%%%%%%%%%%
\begin{figure}
\includegraphics[height=.30\textheight]{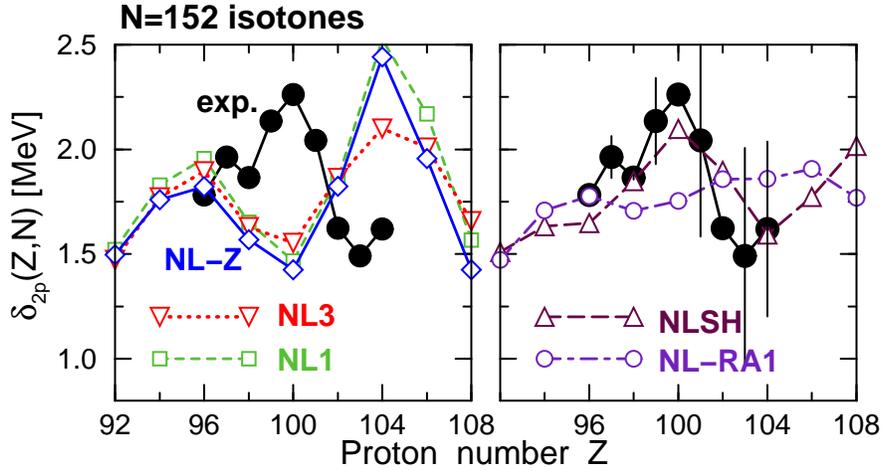}
\caption{The $\delta_{2p}(Z,N)$ quantity for the chain of $N=152$ isotones 
obtained in the CRHB+LN calculations with indicated RMF parametrizations. 
Solid circles are used for experimental data, while open symbols for 
theoretical results. The experimental error bars are shown in right panel.}
\label{delta-n152}
\end{figure}
%%%%%%%%%%%%%%%%%%%%%%  end of figure %%%%%%%%%%%%%%%%%%%%%%%%

 We study the accuracy of the description of shell structure with different parametrizations
of the RMF Lagrangian in the deformed $A\sim 250$ mass region.
Fig.\ \ref{delta-n152} compares experimental and calculated $\delta_{2p}(Z,N)$
quantities for the $N=152$ isotone chain. The experimental data shows a shell
gap at $Z=100$. Only NLSH describes the position of this gap and the $\delta_{2p}(Z,N)$ 
values agree very well. However, the quasi-particle spectra in Ref.\ 
\cite{AKFAL.02} reveal that this gap lies between the wrong bunches of single-particle
states. Calculations with NLSH also indicate  a gap at $Z=108$, which has not been
observed so far. NL-RA1 does not show any deformed gap for $92 \leq Z \leq 108$. NL3, 
NL1 and NL-Z give a shell gap at $Z=104$, in contradiction with experiment. 

  For the Fm $(Z=100)$ isotope chain, NL3 and NL-RA1 (NL1 and NL-Z) produce a gap at 
$N=148$ ($N=148,150$) instead of at $N=152$ as seen in experiment; NLSH does not show 
a clear gap. Many effective interactions not specifically fitted 
to the actinide region encounter similar problems in the description of deformed shell gaps
in the $A\sim 250$ mass region; see for example Ref.\ \cite{BRBRMG.98}).

%%%%%%%%%%%%%%%%%%%%%%%%%%%%%%%%%%%%%%%%%%%%%%%%%%%%%%%%%%%%%%%
\section{Quasiparticle spectra in odd-mass $A\sim 250$ nuclei}
%%%%%%%%%%%%%%%%%%%%%%%%%%%%%%%%%%%%%%%%%%%%%%%%%%%%%%%%%%%%%%%

  The fact that the experimental quadrupole deformations of the nuclei in this 
mass region are very well described in the CRHB+LN calculations (see Ref.\ 
\cite{AKFAL.02} for details) strongly suggests that the discrepancies 
between experimental and calculated $\delta_{2p}(Z,N)$ are due to 
inaccurate deformed single-particle states with 
present RMF parametrizations. This, in turn, is due to errors in the 
positions of spherical subshells from which the deformed
states emerge. Thus, the investigation of the single-particle states in the 
$A\sim 250$ deformed mass region can shed additional light on the reliability of 
the predictions of RMF theory on the energies of spherical subshells  
responsible for 'magic' numbers in superheavy nuclei. This is because several deformed 
single-particle states experimentally observed in odd nuclei of this mass 
region (see Table \ref{qp-experiment}) originate from these subshells.

A proper description of odd nuclei implies the loss of the time-reversal 
symmetry of the mean-field, which is broken by the unpaired nucleon. The 
BCS approximation has to be replaced by the Hartree-(Fock-)Bogoliubov method, 
with time-odd mean fields taken into account. The breaking of time-reversal
symmetry leads to the loss of the double degeneracy (Kramer's degeneracy) 
of the quasiparticle states. This requires the use of the signature or 
simplex basis in numerical calculations, thus doubling the computing task. 
Furthermore, the breaking of the time-reversal symmetry leads to nucleonic 
currents, which causes {\it nuclear magnetism} \cite{KR.89}. The CRHB(+LN) 
theory takes all these effects into account and thus address for 
the first time the question of a fully self-consistent description of 
quasiparticle states in the framework of the RMF theory. 

 The CRHB code \cite{CRHB} has been extended to describe odd and odd-odd nuclei.  
The blocked orbital can be specified either by its dominant main oscillator 
quantum number $N$or by the dominant $\Omega$ quantum number ($\Omega$ is the 
projection of the total angular momentum on the symmetry axis) of the wave 
function, or by combination of both. In addition, it can be specified by 
the particle or hole nature of the blocked orbital.

\begin{figure}
\includegraphics[height=.44\textheight]{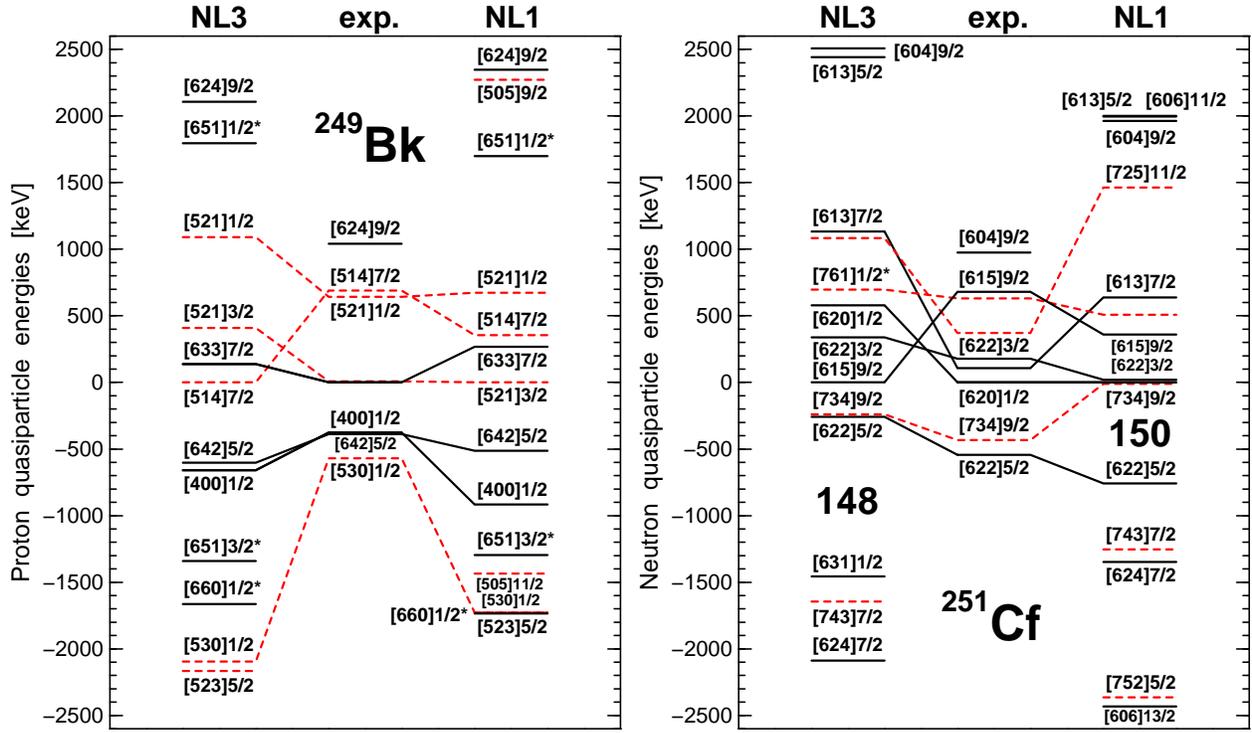}
\caption{Experimental and theoretical quasiparticle energies of neutron 
states in $^{249}$Cf. Positive and negative energies are used for particle 
and hole states, respectively. The experimental data are taken from Ref.\ 
\protect\cite{249Cf}. Solid and dashed lines are used for positive and 
negative parity states, respectively. The symbols 'NL3' and 'NL1' indicate 
the RMF parametrization.  In each 
calculational scheme, attempts were made to obtain solutions for 
every state shown in figure. The absence of a state indicates 
that convergence was not reached.}
\label{odd-nuclei}
\end{figure}

 Experimental and calculated spectra of $^{249}$Bk and $^{251}$Cf are compared 
in Fig.\ \ref{odd-nuclei}. This is the first ever direct comparison between experiment 
and theoretical quasiparticle spectra obtained for deformed nuclei within the 
framework of the RMF theory. The CRHB calculations have been performed with D1S 
force \cite{D1S} in the particle-particle channel and with NL1 and NL3 parametrizations
\cite{NL1,NL3} of the RMF Lagrangian. Since the results are discussed in detail
in Ref.\ \cite{AKFAL.02}, only main features will be outlined below.

%%%%%%%%%%%%%%%%%%%%%%%%%%%%%%%%%%%%%%%%%%%%
\begin{table}
\begin{tabular}{cccc}
\hline
Proton states & & Neutron states & \\ 
Spherical subshell  &   Deformed state        & Spherical subshell  & Deformed state              \\  \hline
                    &                         & $\nu 1k_{17/2}$         &  {\bf $\nu$ [880]1/2}   \\    
$\pi 1j_{15/2}$     &  {\bf $\pi$ [770]1/2}   & $\nu 2h_{11/2}$         &  {\bf $\nu$ [750]1/2}   \\
$\pi 3p_{1/2} $     &    N/A                  & $\nu 1j_{13/2}$         &  $\nu [761]1/2$         \\
$\pi 3p_{3/2} $     &    N/A                  & $N=184        $         &  --------------------   \\
$\pi 1i_{11/2}$     &  {\bf $\pi$ [651]1/2}   & $\nu 4s_{1/2} $         &  N/A                    \\
$Z=120$             &  --------------------   & $\nu 3d_{5/2} $         &  $\nu [620]1/2$         \\
$\pi 2f_{5/2} $     &  $\pi [521]1/2$         & $\nu 3d_{3/2} $         &  N/A                    \\
$\pi 2f_{7/2} $     &  $\pi [521]3/2$,\,\,$\pi [530]1/2$ & $N=172$      &  --------------------   \\
$\pi 1i_{13/2}$     &  $\pi [642]5/2$,\,\,$\pi [633]7/2$,\,\,$\pi [624]9/2$ &
                     $\nu 2g_{7/2} $         &  $\nu [622]3/2$                  \\
$\pi 3s_{1/2} $     &  $\pi [400]1/2$                                       
                    &$\nu 2g_{9/2} $         &   $\nu [622]5/2$, $\nu [613]7/2$, $\nu [604]9/2$ \\
$\pi 1h_{9/2} $     &  $\pi [514]7/2$        & $\nu 1j_{15/2}$  &   $\nu [734]9/2$, $\nu [725]11/2$ \\
                    &                        &  $\nu 1 i_{11/2}$        &   $\nu [615]9/2$, $\nu [624]7/2$ \\
\hline
\end{tabular}
\caption{Spherical subshells active in superheavy nuclei and their deformed 
counterparts active in the $A\sim 250$ mass region. The left column shows the 
spherical subshells active in the vicinity of the ``magic'' spherical gaps 
$(Z=120, N=172)$. Their ordering is given according to the RMF calculations 
with the NL3 parametrizations in the $^{292}_{172}120$ system (see Fig.\ 
\protect\ref{z120-tot}). Although the gaps depend on the specific RMF 
parametrization, the same set of spherical subshells is active with other 
parametrizations (see, for example, Fig.\ 4 in Ref.\ \protect\cite{BRRMG.99}). 
The right column shows the deformed quasiparticle states observed in 
$^{249}_{\,\,\,97}$Bk$_{152}$ \protect\cite{249Bk} and 
$^{249,251}_{\,\,\,98}$Cf$_{151,153}$ \protect\cite{249Cf,251Cf}. 
The bold style is used for the states which might be observed when either 
proton or neutron number is increased by $\approx 10$ as compared with these
nuclei. The symbols 'N/A' (not accessible) are for the deformed states which 
typically increase their energy with increasing deformation and thus are not 
likely to be seen experimentally.}
\label{qp-experiment}
\end{table}
%%%%%%%%%%%%%%%%%%%%%%%%%%%%%%%%%%%%%%%%%%%%%%%%%%%%%%%%%%%%%%%%%%%%%%%%%%%%%

 Although the same set of quasiparticle states as in experiment appears, 
the calculated spectra are less dense. This is related to the effective mass 
(Lorentz mass in the notation of Ref.\ \cite{JM.89}) of the nucleons at the 
Fermi surface $m^*(k_F)/m$. While the experimental density of the quasiparticle 
levels corresponds to $m^*(k_F)/m$ close to one,  the low effective mass 
$m^*(k_F)/m \approx 0.66$ of the RMF theory \cite{BRRMG.99} leads to a 
stretching of the energy scale. It has
been demonstrated for spherical nuclei that the particle-vibration coupling brings the 
average level density in closer agreement with experiment \cite{MBBD.85}.
%, which brings 
% $m^*(k_F)/m$ closer to one. 
In a similar way, the particle-vibration coupling leads 
to a compression of the quasi-particle spectra in deformed nuclei \cite{Malov-private}. 
The surface vibrations are less collective in deformed nuclei than in spherical ones 
because they are more fragmented \cite{Sol-book,BM}. As a consequence, the corrections 
to the energies of quasiparticle states in odd nuclei due to particle-vibration 
coupling are less state-dependent in deformed nuclei. Hence the comparison 
between experimental and mean field single-particle states is less ambiguous 
in deformed nuclei as compared with spherical ones \cite{MBBD.85,BM}, at 
least at low excitation energies, where vibrational admixtures to the wave 
functions are small. Assuming for an estimate that the effective mass just 
stretches the energy scale, one can show that the uncertainty of our estimate 
for the spherical subshell energies derived from the energies of deformed states 
can be kept below 300 keV.
% (see Ref.\ \cite{AKFAL.02} for details).

  Fig.\ \ref{odd-nuclei} shows that the calculated energies of a number of states 
are rather close to experiment. On the other hand, the energies of some states 
and their relative positions deviate substantially from experiment. For example, 
only NL1 gives the correct ground state $\nu [620]1/2$ in $^{251}$Cf, whereas NL3 
gives the $\nu [615]9/2$. Detailed analysis shows that the discrepancies 
between experiment and calculations can be traced back to energies of 
spherical subshells from which deformed states emerge. This allows us to define 
'empirical shifts' to the energies 
of spherical subshells which, if incorporated, will correct the discrepancies 
between calculations and experiment seen for deformed quasiparticle states. 
These `empirical shifts' are shown in Fig.\ \ref{z120-tot} as the energy 
difference between self-consistent and corrected energies of specific 
subshells. It is important to note that these corrections lead to a deformed
$N=152$ shell gap and to a larger $Z=100$ shell gap, 
thus improving the description of the shell structure (for example, the 
$\delta_{2p,n}(Z,N)$ quantities) in the deformed $A\sim 250$ mass region.

%%%%%%%%%%%%%%%%%%%%%%%%%%%%%%%%%%%%  figure  %%%%%%%%%%%%%%%%%%%%%%%%%%%
%%%%%%%%%%%%%%%%%%%%%%%%%%%%%%%%%%%%  end figure %%%%%%%%%%%%%%%%%%%%%%%%

%######################################################
\section{Implications for the study of superheavy nuclei}
%######################################################

%%%%%%%%%%%%%%%%%%%%%%  figure %%%%%%%%%%%%%%%%%%%%%%%%%%%%%%
\begin{figure}
\includegraphics[height=.4\textheight]{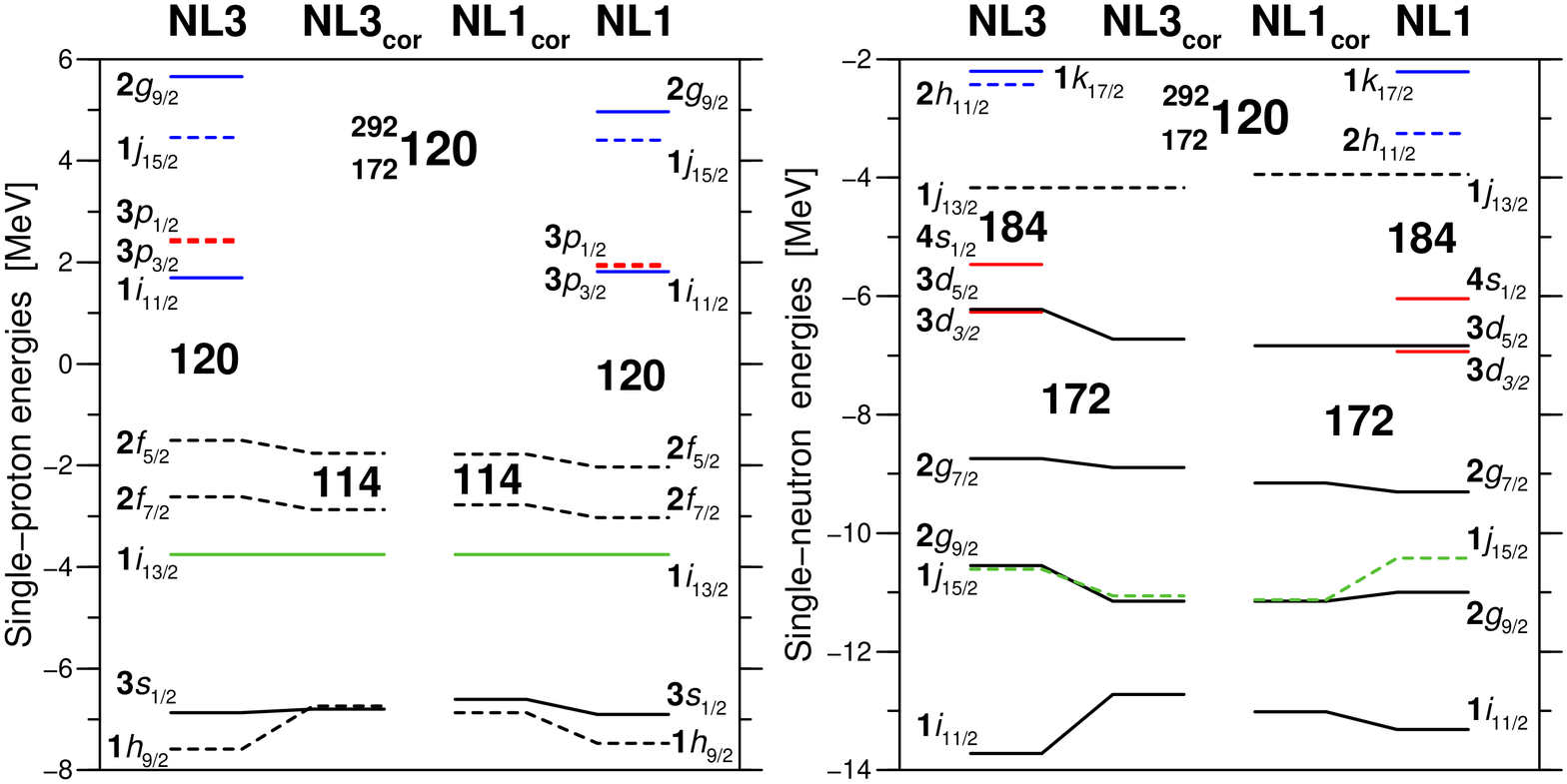}
\caption{Proton  and neutron single-particle states in a 
$^{292}_{172}120$ nucleus. Columns 'NL3' and 'NL1' show the 
states obtained in the RMF calculations at spherical shape 
with the indicated parametrizations. For proton system, the 
energy of the $1i_{13/2}$ state in the NL1 parametrization 
is set to be equal to the one in NL3, which means that the 
energies of all states in NL1 (last column) are increased 
by 0.78 MeV. For neutron sysytem, the energies of all states 
obtained with the NL1 parametrization (last column) are 
increased by 0.76 MeV in order to have the same energies 
of the $2g_{9/2}$ states in the second and third columns.
The columns 'NL3$_{\rm cor}$' and 'NL1$_{\rm cor}$' show how 
the spectra are modified if empirical shifts were introduced 
based on discrepancies between calculations and experiment 
for quasiparticle spectra in $^{249}$Bk and $^{249,251}$Cf.
Solid and dashed lines are used for positive and negative 
parity states. Spherical gaps at $Z=114,Z=120$ and at 
$N=172,N=184$ are indicated.}
\label{z120-tot}
\end{figure}
%%%%%%%%%%%%%%%%%%%%%%  end of figure %%%%%%%%%%%%%%%%%%%%%%%

 In the NL1 and NL3 parametrizations, the energies of the spherical
subshells, from which the deformed states in the vicinity of the Fermi
level of the $A\sim 250$ nuclei emerge, are described with an accuracy
better than 0.5 MeV for most of the subshells (see Fig.\ \ref{z120-tot}
where 'empirical shifts', i.e. corrections, for single-particle energies 
are indicated). The discrepancies (in the range of 0.6-1.0 MeV) are 
larger for the $\pi 1h_{9/2}$ (NL3, NL1), $\nu 1i_{11/2}$ (NL3), 
$\nu 1j_{15/2}$ (NL1) and $\nu 2 g_{9/2}$ (NL3) spherical subshells. 
Considering that the RMF parametrizations were fitted only to bulk 
properties of spherical nuclei this level of agreement is good. The 
NL-Z \cite{NLZ} force provides comparable level of accuracy.

  In contrast, the accuracy of the description of single-particle 
states is unsatisfactory in the NLSH and NL-RA1 parametrizations, 
where 'empirical shifts' to the energies of some spherical subshells are 
much larger than in NL1 and NL3. NL-SH and NL-RA1 are the only RMF 
sets indicating $Z=114$ as a magic proton number \cite{LSRG.96,NL-RA1}.
In the light of present results, this prediction should be treated with 
a considerable caution.

 The spectra of spherical magic superheavy nuclei are not modified
much with empirical shifts (see Fig.\ \ref{z120-tot} for the calculated
and corrected single-particle spectra of a $^{292}_{172}$120 nucleus).
Such a study relies on the assumption that these corrections (which
essentially apply to the $l$-shells from which spin-orbit partner
$j$-shells ($j=l\pm 1/2$) emerge) should be similar in deformed 
$A\sim 250$ mass region and in superheavy nuclei. The corrected spectra 
from the NL1 and NL3 calculations are very similar with minor differences 
coming from the limited amount of information on quasiparticle states used 
in an analysis. More systematic study of quasiparticle states in deformed 
nuclei are required to determine these corrections more precisely.

   Let us consider the calculations for the $Z=120$, $N=172$ nucleus. The corrected 
spectra still suggest that $N=172$ and $N=184$ are candidates for magic neutron 
numbers in superheavy nuclei. The position of the $\nu 4s_{1/2}$ spherical 
subshell and the spin-orbit splitting of the $3d_{5/2}$ and $3d_{3/2}$ subshells 
will decide which of these numbers (or both of them) is (are) magic. The corrected
proton spectra indicate that
the $Z=120$ gap is large whereas the $Z=114$ gap is small. Hence, on the
basis of the present investigation we predict that $Z=120$ is the magic proton
number. This conclusion is based on the assumption that the NL1 and NL3 sets
predict the position of the $\pi 1i_{11/2}$ and $\pi 3p_{1/2,3/2}$ subshells
within 1 MeV error. The positions of $\pi 1 j_{15/2}$ and $\pi 2g_{9/2}$
seems less critical, because they are located well above this group of states
both in Skyrme and RMF calculations \cite{BRRMG.99}. It seems possible to obtain
information about the location of the $\pi 1i_{11/2}$ subshell, which may have
been observed through its deformed state $(\pi [651]1/2)$ in superdeformed
rotational bands of Bi-isotopes \cite{Bi-SD1,Bi-SD2}. An CRHB analysis may
provide this critical information.

  The Nilsson diagrams given, for example, in Figs.\ 3 and 4 of Ref.\
\cite{CAFE.77} suggest that spectroscopic studies of deformed odd nuclei
with proton and neutron numbers up to $Z\approx 108$ and $N\approx 164$
may lead to  observation of the deformed states with $\Omega=1/2$,
emerging from the $\pi 1i_{11/2}$ and $\pi 1j_{15/2}$ spherical subshells
located above the $Z=120$ shell gap and from $\nu 1k_{17/2}$ and either
$\nu 2 h_{11/2}$ or $\nu 1j_{13/2}$ subshells located above the $N=184$
shell gap. This will further constrain microscopic models and effective
interactions.

 No information on low-$j$ states, such as $\pi 3p_{3/2}$, $\pi 3p_{1/2}$,
$\nu 3d_{3/2}$ and $\nu 4s_{1/2}$, which decide whether $Z=120$ or $Z=126$
and $N=172$ or $N=184$ are magic numbers in microscopic theories (see Refs.\
\cite{BRRMG.99,RBM.01} and references quoted therein) will come from the
study of deformed nuclei (see Table \ref{qp-experiment}).

    The measured and calculated energies of the single-particle states at
normal deformation provide constraints on the spherical shell gaps of
superheavy nuclei. In particular, the small splitting between the
$\pi [521]1/2$ and $\pi [521]3/2$ deformed states, from the $\pi 2f_{5/2}$
and $\pi 2f_{7/2}$ spherical subshells that straddle proton number 114,
suggests that the $Z=114$ shell gap is not large. More systematic studies 
of the splitting between the $\pi [521]1/2$ and $\pi [521]3/2$ deformed 
states may provide more stringent information on whether a shell gap 
exists at $Z=114$.

\section{Conclusions}

 The cranked relativistic Hartree+Bogoliubov theory has been applied for
a systematic study of the nuclei around $^{254}$No, the heaviest nuclei
for which detailed spectroscopic data are available. The deformations,
rotational response, pair correlations, quasiparticle spectra, shell
structure and two-nucleon separation energies have been studied.
The part of this study devoted to the investigation of quasiparticle
states and its implications for the study of superheavy nuclei is presented
in this contribution. It is concluded that the energies of the 
spherical subshells, from which active deformed states of these nuclei
emerge, are described with an accuracy better than 0.5 MeV for most of
the subshells with the NL1 and NL3 parametrizations. However, for a 
few subshells the discrepancies reach 0.7-1.0 MeV. Amongst 
the investigated RMF sets,  NL1, NL3 and NL-Z provide best description 
of single-particle states so they are recommended  for the study of 
superheavy nuclei. The corresponding self-consistent calculations
predict as candidates for magic numbers $N=172$ and $N=184$ for neutrons
and $Z=120$ for protons. No significant shell gap is found at $Z=114$. These
conclusions take into account the possible shifts of spherical subshells that
are suggested by the discrepancies between calculations and experiment
for deformed states in the $A\sim 250$ mass region found in our analysis.

%%%%%%%%%%%%%%%%%%%%%%%%%%%%%%%%%%%%%%%%%%%%%%%%
%% BACKMATTER
%%%%%%%%%%%%%%%%%%%%%%%%%%%%%%%%%%%%%%%%%%%%%%%%

\begin{theacknowledgments}
The authors would like to thank P.\ Ring, R.\ R.\ Chasman and A.\ Malov
for valuable discussions. 
% Argonne National Laboratory's work was supported 
% by the U.S. Department of Energy, Office of Science, Office of Basic Energy 
% Sciences, under contract W-31-109-Eng-38. 
This work was supported in part by the U. S. Department of Energy, Nuclear
Physics Division, under Contracts  
No. W-31-109-ENG38 and DE-FG02-95ER40934.
The numerical calculations were made in part on the Cray PVP Cluster at the
National Energy Research Scientific Computing Center.
\end{theacknowledgments}

%% If the bibliography is produced without BibTeX comment out the

\bibliographystyle{aipprocl} % if natbib is missing

\end{document}